\documentclass[11pt,reqno]{amsart}
\newtheorem{lemma}{Lemma}
\newtheorem{theorem}{THEOREM}

\newcommand\N{{\mathbb N}}
\newcommand\C{{\mathbb C}}
\newcommand\R{{\mathbb R}}
\newcommand\Tr{{\rm Tr}}
\newcommand\half{{\mbox{$\frac 12$}}}
\newcommand\eps{\varepsilon}

\newcommand\infspec{{\rm inf\, spec\, }}
\newcommand\id{{\mathbb I}}
\def\an#1{a_{#1}^{\phantom{\dagger}}}
\def\ad#1{a_{#1}^\dagger}
\newcommand\vecz{{\bf z}}

\begin{document}

\title[Rotating Bose Gases]{VORTICES AND SPONTANEOUS SYMMETRY BREAKING
  IN ROTATING BOSE GASES} 

\thanks{Plenary talk given at QMath10, $10^{\rm th}$ Quantum Mathematics
  International Conference, Moeciu, Romania, September 10--15, 2007.
  \\ \indent \copyright\, 2008 by the author. This work may be reproduced, in
  its entirety, for non-commercial purposes.}

\author{ROBERT SEIRINGER}

\address{Department of Physics, Princeton University, Jadwin Hall,
Princeton NJ 08542-0708, USA}
\email{rseiring@princeton.edu}

\begin{abstract}
  We present a rigorous proof of the appearance of quantized vortices
  in dilute trapped Bose gases with repulsive two-body interactions
  subject to rotation, which was obtained recently in joint work with
  Elliott Lieb \cite{LSrot}. Starting from the many-body Schr\"odinger
  equation, we show that the ground state of such gases is, in a
  suitable limit, well described by the nonlinear Gross-Pitaevskii
  equation. In the case of axially symmetric traps, our results show
  that the appearance of quantized vortices causes spontaneous
  symmetry breaking in the ground state.
\end{abstract}

\maketitle

\section{Introduction}

In recent remarkable experiments \cite{abo,hodby,madison,matt}, the
appearance of quantized vortices in the ground state (and low
temperature equilibrium states) of rotating dilute Bose gases was
beautifully demonstrated. These quantized vortices are a consequence
of the superfluid nature of the system under investigation. In
particular, since the system is almost completely Bose condensed, it
behaves like a single quantum particle.

The state of ultracold dilute Bose gases is usually described by means
of the Gross-Pitaevskii (GP)
  equation \cite{BR99,CD99,GP99,FS01,AD01}. This non-linear
Schr\"odinger equation originates as the variational equation from the
corresponding {\it GP energy functional}, given by
\begin{equation}\label{defgpf}
{\mathcal E}^{\rm GP}[\phi] = \langle \phi| H_0 |\phi\rangle 
+ 4\pi g \int_{\R^3} |\phi(x)|^4 d^3\!x \ .
\end{equation}
Here, $\phi \in L^2(\R^3)$, and $H_0$ denotes the one-particle
Hamiltonian, describing the kinetic, potential and rotational energy
of the particles. In fact, if $\Omega $ denotes the angular velocity
vector and $V(x)$ the trap potential, $H_0$ is, in appropriate units,
given by
\begin{equation}\label{defh0}
H_0 = -\Delta + V(x) - \Omega\cdot L \,,
\end{equation}
where $L= -i x \wedge \nabla$ denotes the angular momentum
operator. The parameter $g$ in (\ref{defgpf}) is nonnegative and
measures the interaction strength among the particles. The trap
potential $V(x)$ is assumed to be locally bounded and to increase fast
enough at infinity in order to have the particles confined to the trap
(and, in particular, to ensure that $H_0$ is bounded from below). More
precisely, we assume that
\begin{equation}\label{assv}
\lim_{|x|\to \infty}  \left( V(x) - \frac 14 |\Omega\wedge x|^2 \right) = +\infty\,.
\end{equation}
Since $-\Delta - \Omega\cdot L = (-i\nabla + \Omega \wedge x/2)^2 - |\Omega\wedge
x|^2/4$, this implies the desired property. 

The GP energy is the minimal value of ${\mathcal E}^{\rm GP}[\phi]$
among all (appropriate normalized) functions $\phi$, i.e.,
$$
E^{\rm GP}(g,\Omega)= \inf_{\|\phi\|_2=1} {\mathcal E}^{\rm GP}[\phi]\,.
$$
Using (\ref{assv}) and the fact that $g\geq 0$, it is in fact not
difficult to show that the infimum is actually a minimum (see
\cite{LSY00}). That is, there exists a minimizer of the GP
functional (\ref{defgpf}). Note that, in general, there may be many
different minimizers. In any case, any minimizer satisfies the {\it GP
  equation}
$$
\boxed{ \phantom{\int} 
-\Delta\phi(x) +V(x)\phi(x) - \Omega\cdot L \, \phi(x)  + 8\pi g |\phi(x)|^2 \phi(x) = 
\mu \phi(x)\qquad } 
$$
where $\mu = E^{\rm GP}(g,\Omega) + 4\pi g \int_{\R^3} |\phi(x)|^4 d^3\!x$ is the
corresponding chemical potential.

For axially symmetric $V(x)$, i.e., in case $V(x)$ commutes with
$\Omega\cdot L$, the GP functional is invariant under rotation about
the $\Omega$ axis. It turns out that for any $\Omega\neq 0$, this
rotational symmetry is broken in the GP minimizer for large enough
interaction strength $g$ \cite{S02,S03}. This symmetry breaking is the
result of the appearance of quantized vortices since, in case of more
than one vortex, they cannot be arrange in a symmetric way. Note that,
in particular, this implies that there will be many GP minimizers (for
$g$ large enough).

We remark that the phenomenon just described is a special feature of
rotating systems and cannot be observed in a non-rotating system. In
fact, for $\Omega=0$ there is always a unique minimizer of the GP
functional \cite{LSY00}.

It turns out that the appearance of quantized vortices, and the
resulting symmetry breaking, which we have just described, are not
merely a property of the GP theory, but can actually be derived out of
the underlying (many-particle) Schr\"odinger equation. This was proved
in \cite{LSrot}. In the following sections, we will give a
summary of these results, and we will explain the key ideas leading to
their proof.

\section{The Schr\"odinger Equation for Many Particles}

Consider a quantum-mechanical system of a large number, $N$, of
bosons, with one-particle energies described by $H_0$ (given in
(\ref{defh0}) above). We assume that the particles interact via a
repulsive pair interaction potential $v_a(x)$. The Hamiltonian for
this system is given by
\begin{equation}\label{ham}
H_{N} = \sum_{i=1}^N H_0^{(i)} + \sum_{1\leq i<j\leq N} v_{a}(x_i-x_j)
\,,
\end{equation}
where the superscript $(i)$ refers to the fact that $H_0$ acts on the
$i$'th variable.
Since the particles under consideration are bosons, the Hamiltonian
$H_N$ acts on the subspace of totally symmetric functions in
$\bigotimes^N L^2(\R^3)$, which we denote by ${\mathcal H}_N$.

The interaction potential $v_a(x)$ is assumed to be nonnegative and of
short range. More precisely, it is assumed to have finite {\it scattering
length} \cite{LSY00,LSSY}, denoted by $a$, which means that it has to be integrable at
infinity (i.e., it has to decay faster than $|x|^{-3}$). A typical
example would be a hard sphere interaction, which formally means 
that $v_a(x) =\infty$ for $|x|\leq a$ and $v_a(x)=0$ otherwise. We
shall, in fact, choose some fixed (nonnegative) interaction potential
$w(x)$ with scattering length $1$ and obtain $v_a(x)$ by scaling as
$$
v_{a}(x) = a^{-2} w(x/a) \, .
$$
It is then easy to see that $v_a(x)$ has scattering length
$a$. Moreover, $a$ now appears as a parameter in the Hamiltonian $H_N$,
which can be freely varied. In particular, we can (and will) let $a$
depend on $N$. We note that  this scaling of $v_a(x)$ is, of
course, mathematically and physically equivalent to scaling the trap
potential $V(x)$ (and the angular velocity $\Omega$) in an appropriate
way, while keeping the interaction potential fixed.

\subsection{Ground State Energy}

For fixed $w(x)$ and $V(x)$, we shall denote the ground state energy
of $H_N$ as $E_0(N,a,\Omega)$, i.e.,
$$
E_0(N,a,\Omega) = \inf_{\Psi\in {\mathcal H}_N} \frac{ \langle \psi |
  H_N|\Psi\rangle}{\langle\Psi|\Psi\rangle}\,.
$$
Since the ground state energy per unit volume of a homogeneous
Bose gas with interaction $v_a(x)$ at density $\rho$ is given by $4\pi
a \rho^2$ for low density \cite{LY1}, it is reasonable to expect that $E_0(N,a,\Omega) \approx N
E^{\rm GP}( N a ,\Omega)$ for dilute gases. Here, dilute means that
$a^3\bar\rho \ll 1$, where $\bar\rho$ denotes the average
density. This condition is, in particular, satisfied if $N\gg 1$ and
$N a=O(1)$. We call this the {\it GP limit}. In this limit, we have the
following result \cite{LSrot}.

\begin{theorem}\label{energy}
For any $g\geq 0$ and $\Omega\in \R^3$, 
\begin{equation}\label{eq:thm1}
\boxed{\ \lim_{N\to\infty} \frac {E_0(N,g/N,\Omega)}{N} = E^{\rm
    GP}(g,\Omega) \ }
\end{equation}
\end{theorem}

That is, for large $N$ and $a = O(1/N)$, the ground state energy per particle is
given by the GP energy with coupling parameter $g=Na$.
Theorem~\ref{energy} holds for all angular velocities $\Omega$
(satisfying the stability criterion (\ref{assv})).  It extends
previous results in the nonrotating case $\Omega=0$ \cite{LSY00}.

Note that the right side of (\ref{eq:thm1}) is independent of the
choice of the unscaled interaction potential $w(x)$. In the dilute
limit considered here, only the scattering length $a$ matters, and not
the details of the interaction potential. Note also that the result
cannot be obtained by simple perturbation theory; in fact, the
$\int|\phi|^4$ term in the GP functional is partly kinetic energy, and
not the average value of $v_a(x)$ (which might even be zero, as in the
case of the hard-sphere interaction).

As will be pointed out in Subsect.~\ref{subsec:ags} below, it is
essential to restrict to symmetric wave functions (bosons) in
Theorem~\ref{energy}. For the absolute ground state energy (defined as the
infimum of $H_N$ over all wavefunctions, not necessarily symmetric
ones), the result is wrong, in general. For the absolute ground state,
the right side has to be replaced by minimizing a density-matrix
functional instead \cite{S03}.

\subsection{Bose-Einstein Condensation}

The GP energy functional (\ref{defgpf}) and its minimizers contain
information not only about the ground state energy of the many-body
Hamiltonian (\ref{ham}), but also about the ground state or, more
precisely, its reduced density matrices. Recall that for any
wavefunction $\Psi \in {\mathcal H}_N$, its {\it reduced one-particle
density matrix} $\gamma_N^{(1)}$ is given by the kernel
$$
\gamma_N^{(1)}(x,x') = N \int_{\R^{3(N-1)}} \Psi(x,x_2,\dots,x_N)
\Psi^*(x',x_2,\dots,x_N) d^3\!x_2\cdots d^3\!x_N\,.$$ Note that this
defines a positive trace class operator on the one-particle space $L^2(\R^3)$.

The one-particle density matrix of a state $\Psi$ contains all the
information about the system concerning expectation values of
one-particle operators. It particular, the concept of {\it Bose-Einstein
condensation} (BEC) is defined in terms of $\gamma_N^{(1)}$.

Note that if $\Psi$ is normalized, i.e., $\|\Psi\|_2=1$, then the
trace of $\gamma_N^{(1)}$ is $N$.  BEC means that $\gamma_N^{(1)}$ has
an eigenvalue of order $N$. The corresponding eigenfunction is called
the {\it condensate wave function}. For dilute systems, as we consider here,
one expects in fact {\it complete BEC}, meaning that $\gamma_N^{(1)}$ is
approximately a rank one projection, or $\gamma_N^{(1)}(x,x')\approx N
\phi(x)\phi(x')$ for some normalized $\phi\in L^2(\R^3)$.

In the non-rotating case $\Omega=0$, complete BEC in the ground state
of $H_N$ was proved in \cite{LS}. Moreover, it was
shown that the condensate wave function equals the GP
minimizer. Recall that in the case $\Omega=0$ there is a unique
minimizer of the GP functional (\ref{defgpf}) (up to constant phase
factor, of course), which we denote by $\phi^{\rm GP}$. That is, if
$\gamma_N^{(1)}$ denotes the one-particle density matrix of the ground
state $\Psi$ of $H_N$ for $\Omega=0$, then
\begin{equation}\label{beco0}
\boxed{\ \lim_{N\to \infty}\frac 1N \gamma_N^{(1)}(x,x') = \phi^{\rm
    GP}(x) \phi^{\rm GP}(x')\ }
\end{equation}
in the GP limit $N\to \infty$, $Na\to g$. To be precise, the limit
(\ref{beco0}) holds in trace norm sense. Note that although $a$ is
scaled to zero in the limit considered, the right side of
(\ref{beco0}) depends on $g=Na$ via $\phi^{\rm GP}$.

The corresponding result for $\Omega\neq 0$ is necessarily more
complicated because of non-uniqueness of the GP minimizer $\phi^{\rm
  GP}$. It is actually more natural to not just look at a ground state
of $H_N$ (which may not be unique in the rotating case either), but on
the set of all {\it approximate ground states}. These are defined as
sequences of (bosonic) $N$-particle density matrices $\gamma_{N}$
(that is, positive operators on ${\mathcal H}_N$ with trace one) with
$ \Tr\, H_{N}\gamma_{N} \approx N E^{\rm GP}$. One can then expect
that the reduced one-particle density matrix $\gamma_N^{(1)}$ of any
such approximate ground state is a convex combination of GP
minimizers, i.e.,
$$
\gamma_N^{(1)}(x,x') \approx \sum_i \lambda_i \phi_i^{\rm GP}(x)\phi_i^{\rm
  GP}(x')^*$$
where each $\phi_i^{\rm GP}$ is a GP minimizer, and $\sum_i \lambda_i = N$.

Theorem~\ref{condensation} below states that this is indeed the case. 
The mathematically precise formulation is slightly complicated
by the fact that the set of GP minimizers is, in general, not
countable.

Let $\Gamma$ be the set of all limit points of one-particle density
matrices of approximate ground states:
\begin{align}\nonumber
  \Gamma &= \left\{ \gamma\, : \, \exists {\rm \ sequence\ } \gamma_{N},
     \lim_{N\to\infty, \, Na\to g} \frac 1N \Tr\, H_{N} \gamma_{N}=
    E^{\rm GP}(g,\Omega), \right. \\ & \label{defgamma} \qquad\qquad \left.
    \lim_{N\to\infty}\frac 1N \gamma_{N}^{(1)}=\gamma\right\}\,.
\end{align}
Since $H_0$ has a compact resolvent by our assumption (\ref{assv}),
one easily sees that $\Tr\,\gamma=1$ for all $\gamma\in
\Gamma$. Moreover, because of the linearity of the conditions in
(\ref{defgamma}), $\Gamma$ is clearly convex.

\begin{theorem} \label{condensation} For given value of $g\geq 0$ and
  $\Omega$, let $\Gamma$ denote the set of all limit points of
  one-particle density matrices of approximate ground states of $H_N$, defined in
  (\ref{defgamma}).
\begin{itemize}
\item[(i)] $\Gamma$ is a compact and convex subset of the set of all
  trace class operators.
\item[(ii)] Let $\Gamma_{\rm ext}\subset \Gamma$ denote the set of
  extreme points in $\Gamma$.  We have $\Gamma_{\rm
    ext} = \{ |\phi\rangle\langle\phi|\, : \, {\mathcal E}^{\rm GP}[\phi]=E^{\rm
    GP}(g,\Omega)\}$, i.e., the extreme points in $\Gamma$ are given by the
  rank-one projections onto GP minimizers.
\item[(iii)] For each $\gamma\in \Gamma$, there is a positive (regular Borel) measure
  $d\mu_\gamma$, supported in $\Gamma_{\rm ext}$, with
  $\int_{\Gamma_{\rm ext}} d\mu_\gamma(\phi) =1$, such that
$$
\boxed{\ 
\gamma = \int_{\Gamma_{\rm ext}} d\mu_\gamma(\phi)\,
|\phi\rangle\langle\phi| \ }
$$
where the integral is understood in the weak sense. 
That is, every
$\gamma\in\Gamma$ is a convex combination of rank-one projections onto
GP minimizers.
\end{itemize}
\end{theorem}

We remark that item (iii) of Theorem~\ref{condensation} follows from
item (ii) by Choquet's Theorem \cite{choq}.

As explained above, Theorem~\ref{condensation} is the natural analogue
of (\ref{beco0}) in the rotating case. It can also be interpreted as a
rigorous proof of superfluidity. As typical for superfluids, angular
momentum in rotating systems is acquired in terms of quantized
vortices. These can be seen by solving the GP equation.

Theorem~\ref{condensation} also shows the occurrence of spontaneous
symmetry breaking. As remarked earlier, axial symmetry of the trap
$V(x)$ leads to non-uniqueness of the GP minimizer for $g$ large
enough \cite{S02,S03}.  Uniqueness can be restored by perturbing $H_0$
to break the symmetry and favor one of the minimizers.  This then
leads to complete BEC in the usual sense, since $\Gamma$ contains
contains only one element in case the GP functional (\ref{defgpf}) has a
unique minimizer.

As in the case of the ground state energy discussed in the previous
subsection, the situation is very different for the absolute ground
state.  The set $\Gamma$ consists of only one element in this case
(namely the minimizer of the density matrix functional discussed
below, which is unique for any value of $\Omega$ and $g$).  In
particular, there is no spontaneous symmetry breaking in the absolute
ground state. This will be discussed in the next subsection.

\subsection{The Absolute Ground State}\label{subsec:ags}

Let $E_{\rm abs}(N,a,\Omega)$ denote the absolute ground state energy
of $H_N$ in (\ref{ham}), irrespective of symmetry constraints, i.e.,
$$
E_{\rm abs}(N,a,\Omega)= \inf_{\Psi\in L^2(\R^{3N})} \frac{ \langle \psi |
  H_N|\Psi\rangle}{\langle\Psi|\Psi\rangle}\,.
$$
Note that necessarily $E_{\rm abs}(N,a,\Omega)\leq
E_0(N,a,\Omega)$. As is well known, for $\Omega=0$ the two energies
are equal. This turns out not to be the case for $\Omega\neq 0$, in
general.

In the GP limit, the absolute ground state energy, and the
corresponding one-particle reduced density matrices of approximate
ground states, turn out to be described by a {\it GP density matrix
functional}, introduced in \cite{S02},
 $${\mathcal E}^{\rm
   DM}[\gamma]={\rm Tr}\left[H_0\gamma\right] +4\pi g
 \int_{\R^3}\rho_\gamma(x)^2d^3\!x \,.$$ Here, $\gamma$ is a positive trace
 class operator on $L^2(\R^3)$, and $\rho_\gamma$ denotes the density
 of $\gamma$, i.e., $\rho_\gamma(x)=\gamma(x,x)$. The functional
 ${\mathcal E}^{\rm DM}$ can be shown \cite{S02} to have a unique
 minimizer (under the normalization condition $\Tr\,\gamma=1$), which
 we denote by $\gamma^{\rm DM}$. We denote the corresponding energy by
 $E^{\rm DM}(g,\Omega)={\mathcal E}^{\rm DM}[\gamma^{\rm DM}]$.

 The following Theorem concerning the absolute ground state of $H_N$ was
 proved in \cite{S03}.

\begin{theorem}\label{abs} For any fixed $g\geq 0$ and $\Omega\in\R^3$, 
$$\boxed{ \ \lim_{N\to\infty}\frac {E_{\rm abs}
(N,g/N,\Omega)} N = E^{\rm DM}(g,\Omega) \quad {\rm and}\quad 
 \lim_{N\to\infty}\frac 1N\gamma_{\rm abs}^{(1)} =\gamma^{\rm
DM}\ }$$
\end{theorem}

Here, $\gamma_{\rm abs}^{(1)}$ denotes the one-particle density matrix
of any approximate (absolute) ground state sequence of $H_N$. In other words, the
set $\Gamma$ defined as in (\ref{defgamma}), but for the absolute
ground state, contains only one element, namely the unique minimizer
of ${\mathcal E}^{\rm DM}$.

Note that ${\mathcal E}^{\rm GP}$ is the restriction of ${\mathcal
  E}^{\rm DM}$ to rank one projections.  In the case of symmetry
breaking (i.e., for $g$ large enough), ${\rm rank\, }\gamma^{\rm
  DM}\geq 2$, and hence $E^{\rm DM}< E^{\rm GP}$. In particular, in
view of Theorems~\ref{energy}--\ref{abs}, the absolute and bosonic
ground state differ significantly, in general, both in terms of their
energy and their reduced one-particle density matrix.

We remark that the results explained in this subsection become
physically relevant if one considers bosons with internal degrees of
freedom. Internal degrees of freedom effectively increase the number
of allowed symmetry classes (see, e.g., \cite{eisen}). In
particular, if the number of states of the internal degrees of freedom
of the bosons is greater or equal to the rank of $\gamma^{\rm DM}$,
$E^{\rm DM}(g,\Omega)$ equals the (bosonic) ground state energy per
particle in the GP limit. More generally, one can show that in the GP
limit the functional ${\mathcal E}^{\rm DM}$, when restricted to
density matrices of rank at most $n$, correctly describes the ground
state energy (and corresponding one-particle density matrix) of bosons
with $n$ internal states.

\section{Sketch of the Proof of Theorem~\ref{energy}}

In the following, we shall give a brief outline of the main ideas in
the proof of Theorem~\ref{energy}. For details we refer to the
original work in \cite{LSrot}. We shall restrict our attention to the
appropriate lower bound on the ground state energy
$E_0(N,a,\Omega)$. The corresponding upper bound can be obtained via a
variational argument, as explained in \cite{S03}.
  
A convenient way to keep track of the bosonic symmetry requirement is
to work in Fock space. Recall that the bosonic Fock space ${\mathcal
  F}$ is given by ${\mathcal F}=\bigoplus_{N\geq 0} {\mathcal H}_N$. In terms
of creation and annihilation operators $\ad{j}$ and $\an{j}$ on
${\mathcal F}$, the Hamiltonian can be written as
\begin{equation}\label{ham2}
H = \sum_{j\geq 1} e_j \ad{j} \an{j} + \half \sum_{ijkl}
\ad{i}\ad{j}\an{k}\an{l} W_{ijkl}\,.
\end{equation}
Here, we choose the basis in the one-particle space $L^2(\R^3)$ as to
diagonalize $H_0$, i.e., $H_0=\sum_j e_j |\varphi_j\rangle\langle
\varphi_j|$, and $\ad{j}$ creates a particle with wavefunction
$\varphi_j$, whereas $\an{j}$ annihilates it. The coefficients
$W_{ijkl}$ are given in terms of expectation values of $v_a(x)$, namely
$W_{ijkl}= \langle \varphi_i\otimes \varphi_j| v_a | \varphi_k\otimes
\varphi_l\rangle$. 

Note that $H$ in (\ref{ham2}) commutes with total particle number
operator $\sum_{j\geq 1} \ad{j}\an{j}$. Hence it splits into a direct sum of
operators on ${\mathcal H}_N$ for $N=0,1,\dots$. In fact, our $H_N$ in
(\ref{ham}) is just the restriction of $H$ to ${\mathcal H_N}$.

The analysis employed for obtaining a lower bound on the ground
state energy of $H$ in the sector of $N$ particles consists of two
main steps:

\begin{itemize}
\item [1.] Eq. (\ref{ham2}) is not necessarily well defined. E.g., if
  $v_a(x)$ is the hard-core interaction potential (or, more generally,
  is not integrable), then $W_{ijkl}= \infty$ for any set of indices.
  In order to overcome this problem, we shall first show that, for a
  lower bound, one can replace $v_a(x)$ by a \lq\lq soft\rq\rq\ and
  longer ranged potential $U(x)$ (with the same scattering length), at
  the expense of the high-momentum part of the kinetic energy. We note
  that this step is necessary even in the case when $v_a(x)$ is
  integrable (and hence (\ref{ham2}) is well defined) in order to
  proceed with the second step.
  
\item [2.] After having replaced $v_a(x)$ by the softer potential $U(x)$,
  one then shows that it is possible to replace the operators $\ad{j}$ and
  $\an{j}$ by complex numbers $z_j$ without changing the ground state
  energy too much \cite{LSYjust}. Note that if all the $\ad{j}$ and $\an{j}$ in
  (\ref{ham2}) are treated as numbers, the expression (\ref{ham2})
  looks very similar to the GP energy functional (\ref{defgpf}); in
  fact, it is given by
$$
\langle \phi_\vecz|H_0|\phi_\vecz \rangle + \half \int_{\R^6} v_a(x-y)
|\phi_\vecz(x)|^2 |\phi_\vecz(y)|^2 \,d^3\!x\, d^3\!y \,,
$$
with $\phi_\vecz(x)= \sum_j z_j \varphi_j(x)$.

\end{itemize}
In the following, we shall explain these two main steps in more detail.

\subsection{Step 1: Generalized Dyson Lemma}

The following Lemma can be viewed as a generalization of an idea of
Dyson \cite{dyson}. The purpose of the lemma is give a lower bound on
the interaction potential $v_a(x)$ in terms of a softer and longer
ranged potential $U(x)$, at the expense of some kinetic energy (see
also \cite{LY1}). For our purpose, we can only spare the high
momentum part of the kinetic energy, however; the low momentum part is
 needed for the $H_0$ term in the GP functional.

We thus have to separate the high momentum from the low momentum part
of the kinetic energy. This can be done in the following way. The
proof of Lemma~\ref{lem:dyson} is given in \cite{LSS}.

\begin{lemma}\label{lem:dyson}
  Let $v_a(x)$ have scattering length $a$ and range $R_0$.  Let
  $\theta_R$ be the characteristic function of the ball $\{x\, : \,
  |x|<R\}$.  Let $0\leq \chi (p) \leq 1$, such that $h(x)\equiv
  \widehat{1-\chi}(x)$ is bounded and integrable,
  $$
  f_R(x)= \sup_{|y|\leq R} | h(x-y) - h(x) |\,,
$$
and
$$
  w_R(x)= \frac{2}{\pi^2} f_R(x) \int_{\R^3} f_R(y)d^3\!y\,.
  $$
  Then for any $\eps>0$ and any positive radial function $U(x)$
  supported in $R_0\leq|x|\leq R$ with $\int U=4\pi$ we have the
  operator inequality
  \begin{equation}\label{eq:lem1}
  -\nabla \chi(p) \theta_R(x) \chi(p) \nabla + \half v_a(x) \geq
  (1-\eps) a U(x) - \frac {a}\eps w_R(x)\,.
  \end{equation}
\end{lemma}

Here, $\chi(p)$ denotes a multiplication operator in momentum space. 
Note that the operator $-\nabla \chi(p) \theta_R(x) \chi(p) \nabla$
can be interpreted as a Laplacian that has been localized to the ball
of radius $R$ and cut off in momentum space. Because of the cut-off,
this is not a local operator, however. The parameter $R$ is chosen
such that $a\ll R\ll N^{-1/3}$. Note that to leading order in $a/R$,
the scattering length of $2aU(x)$ is given in terms of its first
order Born approximation as $(8\pi)^{-1} a \int_{\R^3} U(x)d^3\!x = a$.

Because of the appearance of the characteristic function $\theta_R(x)$
in (\ref{eq:lem1}), Lemma~\ref{lem:dyson} has the following immediate
consequence. If $y_1,\dots, y_n$ are $n$ points in $\R^3$ whose mutual
distance is at least $2R$, then
  \begin{equation}\nonumber
  -\nabla \chi(p)^2 \nabla + \half \sum_{i=1}^n v_a(x-y_i) \geq \sum_{i=1}^n \left[ 
  (1-\eps) a  U(x-y_i) - \frac {a}\eps w_R(x-y_i)\right]\,.
  \end{equation}
  This bound accomplishes the replacement of the hard interaction potential
  $v_a(x)$ by a soft one, at the expense of the high momentum part of the
  kinetic energy.  For given configuration of $N-1$ particles, this
  estimate is  applied to the remaining particle. Of course one
  still has to estimate the contribution from configurations where $2$ (or more)
  of the $N-1$ fixed particles are closer together than $2R$. This can
  be achieved by a Feynman-Kac integral representation \cite{simon} of the ground
  state. We refer to \cite{LSrot} for details.

\subsection{Step 2: Coherent States}

The Fock space $\mathcal F$ can be viewed as an infinite tensor product of the
form ${\mathcal F}=\bigotimes_{j\geq 1} {\mathcal F}_j$, with ${\mathcal F}_j$
spanned by the vectors $(\ad{j})^n|0\rangle$ for $n=0,1,\dots$. Here,
$|0\rangle$ denotes the Fock space vacuum.

Consider first the case of a single mode, ${\mathcal F}_1$, say. For
$z\in \C$, a {\it coherent state} \cite{Klauder} in ${\mathcal F}_1$ is defined by
$$
|z \rangle = e^{-|z|^2/2+z \ad{1}} |0 \rangle \,.
$$ 
These states span in the whole space ${\mathcal F}_1$. In fact, they
satisfy the completeness relation
\begin{equation}\label{compl}
\int_\C dz |z\rangle\langle z|= \id \,,
\end{equation}
where $dz$ stands for $\pi^{-1} dx dy$, and $z=x+iy$, $x,y\in\R$.

In terms of coherent states, {\it upper and lower symbols} of operators can
be defined. Lower symbols are simply the expectation values of
operators in coherent states, e.g., $\langle z|\an{1}|z\rangle = z$
and $\langle z| \ad{1}\an{1}| z\rangle = |z|^2$. Upper symbols, on the
other hand, represent functions of $z$ which, when integrated against
$|z\rangle\langle z| dz$ over $\C$, yield given operators.  For
instance, it is not difficult to see that $\an{1}= \int dz \, z
|z\rangle\langle z|$, while $\ad{1}\an{1}= \int dz \, (|z|^2-1)
|z\rangle\langle z|$. Hence, upper and lower symbols of $\an{1}$ are
given by $z$, whereas the lower symbol of $\ad{1}\an{1}$ is $|z|^2$
and the upper symbol is $|z|^2-1$.

Note that lower symbols yield upper bounds on ground state energies,
by the variational principle, while upper symbols are useful for lower
bounds. The difference in the symbols thus quantifies the error one
makes in replacing the operators $\ad{1}$ and $\an{1}$ by numbers. In
particular, for every quadratic term $\ad{1}\an{1}$ a factor $-1$ has
to be taken into account. For this reason, one cannot introduce
coherent states of all the modes $j$, but only for a finite number of
them. 

In fact, we shall introduce coherent states of all the modes $1\leq
j\leq J$ for some $J\gg 1$.  That is, we first write ${\mathcal
  F}={\mathcal F}_{<}\otimes {\mathcal F}_{>}$, where ${\mathcal F}_<$
is spanned by the vectors of the form $(\ad{1})^{n_1}\cdots
(\ad{J})^{n_J} |0\rangle$, with $n_j\in \N$ for $1\leq j \leq J$.  For
$\vecz=(z_1,\dots,z_J)\in \C^J$, we introduce the projection operator
$\Pi(\vecz)$ on ${\mathcal F}_<$, given by
$$
\Pi(\vecz)=|z_1\otimes \cdots \otimes z_J\rangle\langle z_1\otimes
\cdots \otimes z_J|\,.
$$
Using upper symbols, we can then write the Hamiltonian $H$ in (\ref{ham2}) as
$$
H= \int_{\C^J} d\vecz \, \Pi(\vecz) \otimes h(\vecz)\,.
$$
Here, $h(\vecz)$ represents the upper symbol of $H$. Since only the
modes $1\leq j\leq J$ have been replaced by numbers, $h(\vecz)$ is an
{\it operator} on ${\mathcal F}_{>}$. Using the completeness property
of the coherent states, Eq. (\ref{compl}), it is then easy to see that
$$
\infspec H\geq \inf_{\vecz} \infspec h(\vecz)\,.
$$

One then proceeds to show that $h(\vecz) \approx {\mathcal E}^{\rm
  GP}[\phi_\vecz]$ modulo controllable error terms. These error terms
are, in fact, operators on ${\mathcal F}_{>}$ which describe both the
interactions among particles in high modes as well as the interaction
between particles in modes $j\leq J$ and $j>J$. Precise  bounds on
these terms can be found in \cite{LSrot}.

\section{Sketch of the Proof of Theorem~\ref{condensation}}

In order to obtain information on (approximate) ground states from
bounds on the energy, one proceeds as follows. One first perturbs the
Hamiltonian $H_N$ in (\ref{ham}) by some one-particle perturbation
$S$, and applies the same perturbation to the GP functional
(\ref{defgpf}). One then shows that the result of Theorem~\ref{energy}
still holds for the perturbed system. In fact, the proof of
Theorem~\ref{energy} outlined in the previous section is sufficiently
robust in order to easily incorporate such a modification.

Griffiths' argument \cite{Griffiths} then implies that, for any
$\gamma\in \Gamma$, and any bounded hermitian operator $S$,
\begin{equation}\label{key}
  \Tr\, S\gamma \geq \min_{\phi=\phi^{\rm GP}} \langle \phi|S|\phi\rangle\,,
\end{equation}
where the minimum on the right side is taken over all GP
minimizers. Inequality (\ref{key}) is the key to the proof of
Theorem~\ref{condensation}. The rest follows from convexity
theory \cite{rockafellar}, as we shall explain now.

Recall that an exposed point of a convex set $\mathcal{C}$ is an
extreme point $p$ with the additional property that there is a tangent
plane to $\mathcal{C}$ containing $p$ but no other point of
$\mathcal{C}$.  Hence, for $\widetilde\gamma\in \Gamma$ an exposed
point, there exists an $S$ such that
\begin{equation}\label{tran}
  \Tr\, S \widetilde\gamma \leq  \Tr\, S \gamma   \quad {\rm for\ all\ }\gamma\in\Gamma\,.
\end{equation}
with equality {\it if and only if} $\gamma=\widetilde\gamma$. 

It is not very difficult to show that $|\phi^{\rm GP}\rangle\langle
\phi^{\rm GP}| \in \Gamma$ for any GP minimizer $\phi^{\rm
  GP}$. Hence, if we choose $\gamma$ in (\ref{tran}) to be equal to
$|\phi^{\rm GP}\rangle\langle \phi^{\rm GP}|$ for the $\phi^{\rm GP}$
that minimizes the right side of (\ref{key}) for this particular $S$,
the inequalities (\ref{key}) and (\ref{tran}) imply that 
$$
\min_{\phi=\phi^{\rm GP}} \langle \phi|S|\phi\rangle = \langle
\phi^{\rm GP}|S|\phi^{\rm GP}\rangle \leq \Tr\, S\widetilde\gamma \leq
\Tr\, S \gamma = \langle \phi^{\rm GP}|S|\phi^{\rm GP}\rangle
$$
and hence there is actually equality in (\ref{tran}). This, in turn,
implies that $\widetilde \gamma = |\phi^{\rm GP}\rangle\langle
\phi^{\rm GP}|$. We have thus shown that all exposed points of
$\Gamma$ are of this form!

In order to extend this result to all extreme points, now merely
exposed points, we employ Straszewicz's Theorem \cite{rockafellar},
which states that the exposed points are a dense subset of the extreme
points. Strictly speaking, this theorem only holds in finite
dimensions and not, {\it a priori}, in the infinite dimensional case
under consideration here. However, because of compactness, the set
$\Gamma$ is \lq\lq almost\rq\rq\ finite dimensional, and hence the
theorem can be applied via an approximation argument. We refer again
to \cite{LSrot} for details.

\section{Conclusions}

We have presented a rigorous justification of the Gross-Pitaevskii
approximation for sufficiently dilute rotating Bose gases. For large
particle number $N$ and both $Na$ and $\Omega$ of order 1, the ground
state of a rotating Bose gas is well approximated by the solution to
the GP equation. This is true both for the energy and the reduced
density matrices. In particular, our analysis proves the appearance of
quantized vortices and the occurrence of spontaneous symmetry breaking
in the parameter regime where these phenomena can be observed in the
GP equation, e.g., for $\Omega\neq 0$ and $g$ large enough.

We point out that one of the major open problems in this field is the
validity of the GP equation for {\it rapidly} rotating gases, where
either $|\Omega|\to \infty$ as $N\to \infty$ (in case the trap
potential grows faster than quadratic at infinity), or $\Omega$
approaches the trap frequency (for traps that are asymptotically
quadratic). There is evidence that the GP descriptions breaks down
once the number of vortices in the system is of the same order as the
number of particles. Despite recent progress in this
direction \cite{Y}, a proof of this assertion is still lacking.

\section*{Acknowledgments}
Partial support by U.S. National Science grant PHY 0652356 and by an
A.P. Sloan Fellowship is gratefully acknowledged.

\end{document}